\documentclass[iop]{emulateapj}

\usepackage{graphicx}			
\usepackage{natbib}				
\usepackage{amsmath}			
\usepackage[backref,breaklinks,colorlinks,citecolor=blue]{hyperref}		
	
\usepackage[all]{hypcap}			
\usepackage{cleveref}			
\usepackage{indentfirst}	
\usepackage{fancyhdr}			
\usepackage{amssymb}			
\usepackage{rotating}			

\defcitealias{Maury11}{M11}
\defcitealias{Teixeira12}{T12}
\defcitealias{Plunkett15}{P15}
\defcitealias{OrtizLeon15}{OL15}

\begin{document}


\title{Radio Properties of Young Stellar Objects in the Core of the Serpens South Infrared Dark Cloud}

\author{
Nicholas S. Kern\altaffilmark{1,6},
Jared A. Keown\altaffilmark{2,6},
John J. Tobin\altaffilmark{3},
Adrian Mead\altaffilmark{4,6}
\& Robert A. Gutermuth\altaffilmark{5}
}
\altaffiltext{1}{Department of Astronomy, University of California, Berkeley, CA, USA; nkern@berkeley.edu}
\altaffiltext{2}{Department of Physics and Astronomy, University of Victoria, Victoria, BC, Canada}
\altaffiltext{3}{Leiden Observatory, Leiden University, Leiden, The Netherlands}
\altaffiltext{4}{Department of Astronomy, University of Virginia, Charlottesville, VA, USA}
\altaffiltext{5}{Department of Astronomy, University of Massachusetts, Amherst, MA, USA}
\altaffiltext{6}{Previous summer student at the National Radio Astronomy Observatory}

\submitted{Submitted to AJ}

\begin{abstract}
We present deep radio continuum observations of the star-forming core of the Serpens South Infrared Dark Cloud with the Karl G. Jansky Very Large Array (VLA). Observations were conducted in two bands centered at 7.25 GHz (4.14 cm) and 4.75 GHz (6.31 cm) with a $\sigma_{rms}$ of 8.5 and 11.1 $\mu$Jy/beam, respectively. We also use 2MASS, \emph{Spitzer} and \emph{Herschel} data to put our radio observations in the context of young stellar populations characterized by near and far infrared observations. Within a 5' x 5' region of interest around the central cluster, we detect roughly eighteen radio sources, seven of which we determine are protostellar in nature due to their radio spectral indices and their association with infrared sources. We find evidence for a previously undetected embedded Class 0 protostar and reaffirm Class 0 protostellar classifications determined by previous millimeter wavelength continuum studies. We use our infrared data to derive mid-infrared luminosities for three of our protostellar sources and find relative agreement between the known YSO radio luminosity vs bolometric luminosity correlation. Lastly, we marginally detect an additional six radio sources at the 2-3$\sigma$ level that lie within two arcseconds of infrared YSO candidates, providing motivation for higher sensitivity studies to clarify the nature of these sources and further probe embedded and/or low luminosity YSOs in Serpens South.
\end{abstract}

\keywords{}

\maketitle
\newpage


\section{Introduction}
\label{sec:introduction}
	
	Serpens South is a young stellar cluster that is a part of the broader Aquila Rift complex of dark clouds. Discovered in 2008 by \citet{Gutermuth08} as a part of the \emph{Spitzer} Space Telescope's Gould Belt Legacy Survey, Serpens South has been found to harbor an unusually high ratio of Class I to Class II young stellar objects (YSO), where a Class I YSO is representative of the late mass accretion phase onto a central protostar and Class II of a classical T Tauri pre-main sequence star \citep[see][]{Greene94}. This suggests that Serpens South is in a very early phase of cluster formation and makes it one of the most active sites of star formation within 1 kpc. Since its discovery, it has become the center of a wide range of scholarship. This has consisted of near, mid and far infrared mappings with \emph{Spitzer} and \emph{Herschel} tracing heated dust around protostars \citep{Gutermuth08, Bontemps10}, millimeter mappings tracing cold dust \citep{Maury11}, near infrared polarimetry revealing the importance of global magnetic fields in the cluster's formation history \citep{Sugitani11}, molecular outflows studies \citep[e.g.][]{Nakamura11, Teixeira12, Plunkett15, Plunkett15b}, and a wealth of spectral line surveys probing filamentary infall \citep[e.g.][]{Kirk13, Friesen13, Tanaka13, Fernandez-Lopez14, Nakamura14a}. In spite of the wealth of optical, infrared and submillimeter data currently available, Serpens South has received little attention in the radio continuum. Presently, only one radio continuum study of Serpens South has been conducted by \citet{OrtizLeon15}, who did not detect any radio sources associated with known YSOs in the central core of Serpens South.
	
	Because the core of Serpens South has been shown to harbor a high density of YSOs in the earliest phases of their development, it is an interesting region to search for highly embedded YSOs and allows for a large number of possible protostellar radio detections with only one telescope pointing. The Karl G. Jansky Very Large Array (VLA) is an ideal instrument for this aim. It has been a proven tool for detecting radio emission around YSOs since the early 1990s and its recently upgraded capabilities make it an even more powerful tool for this purpose \citep[e.g.,][]{Curiel89, Anglada98, Beltran01, Shirley07, Dzib13}. The goal of this paper is to for the first time shed light on the radio properties of the clustered protostars in the core of Serpens South.
	
	The study of radio emission around YSOs is an important asset to star formation studies because radio emission can penetrate the high column densities that obscure YSOs at optical and sometimes infrared wavelengths. In the case of highly embedded and very young objects, radio emission can provide evidence for the presence of a central source \citep{Andre00}. Recent studies at millimeter wavelengths have also shown that embedded, low-luminosity protostars can go undetected and overlooked in near infrared imaging of starless cores \citep[e.g.][]{Schnee12}, making long-wavelength studies of star forming regions essential to fully understanding prestellar populations. Additionally, the shape of a YSOs spectral energy distribution (SED) at radio wavelengths can provide information about collimated outflows, magnetic field activity and other high-energy processes around a protostar \citep{Feigelson99}.

	The distances to Serpens South, W40 and the Serpens Main cloud are not agreed upon in the literature. When Serpens South was discovered in 2008, \citet{Gutermuth08} adopted a distance of 260 pc $\pm$ 37 pc based on evidence that its LSR velocities matched LSR velocities of the Serpens Main cloud, which was then thought to be a part of the larger Aquila Rift complex estimated to lie at 260 pc \citep{Straizys03}. However, VLBA parallax measurements conducted in 2010 have established the distance to the Serpens Main cloud as 429 $\pm$ 2 pc \citep{Dzib11}. \citet{Gutermuth08} also argued that Serpens South lies in front of W40, claiming that its cold filaments are seen in absorption against emission from W40. If we are to follow the initial LSR velocity analysis by \citet{Gutermuth08}, we would equate Serpens South to Serpens Main and say Serpens South lies at approximately 429 pc, while W40 lies further away. Indeed, previous radio and x-ray studies adopt a distance of 600 pc to W40, although they admit the distance is poorly constrained \citep{Kuhn10, Rodriguez10}. Here, we adopt a distance of 429 pc to Serpens South, similar to other recent studies of the region \citep{Plunkett15,OrtizLeon15}.

\section{Observations and Data}
\label{sec:obs_and_data}

\begin{figure*}
\label{fig:serpsouth_fov}
\centering
\includegraphics[keepaspectratio=True,scale=.6]{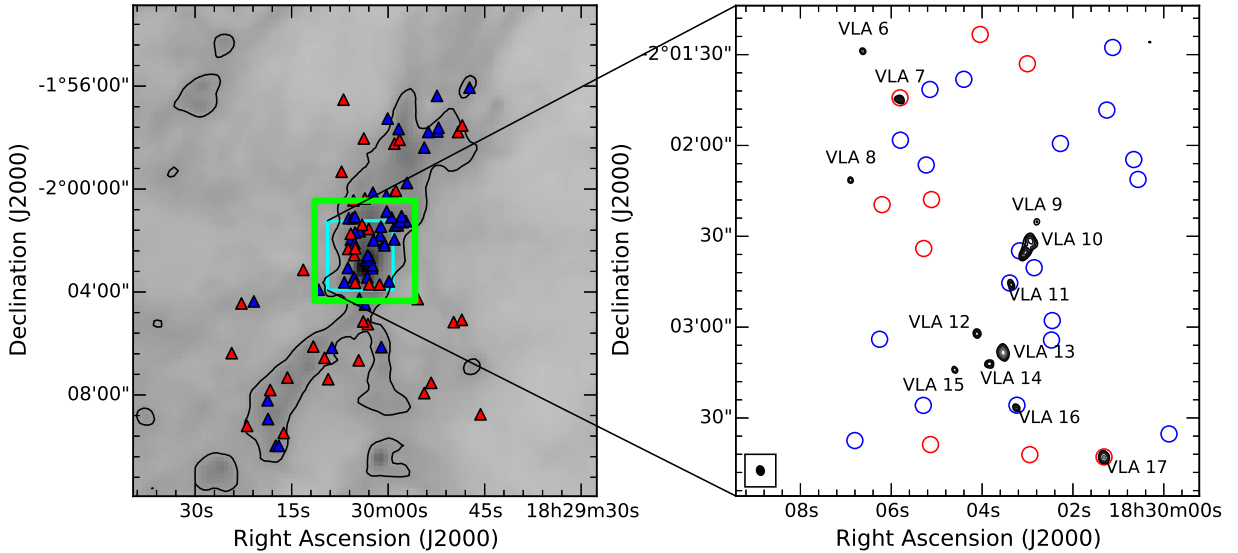}
\caption{\small{{\bfseries Left:} \emph{Herschel} SPIRE 350 $\mu$m greyscale image of Serpens South highlighting its filamentary structure. The black contour intensity level is set at 365 mJy/sr. Blue and red triangles indicate \emph{Spitzer} identified Class I and Class II YSOs respectively (Gutermuth et al. in prep.). The green box indicates our 5 arcmin by 5 arcmin region of interest with the VLA, while the cyan box shows the extent of the zoomed-in figure to the right. {\bfseries Right:} VLA 7.25 GHz radio continuum of the core of Serpens South. The circles represent \emph{Spitzer} Class I and II YSOs.}}
\end{figure*}

We observed Serpens South with one pointing on July 2, 2013, for 1 hour with the VLA in its C array configuration under project number DEM0009. In order to derive a more accurate spectral index, we configured our C band observation to have subbands centered at 4.75 GHz (6.31 cm) and 7.25 GHz (4.14 cm), each with bandwidths of 1.024 GHz. At 4.75 and 7.25 GHz, our images had primary beam diameters of 9.5 and 6.2 arcminutes respectively. Our observation focused on a 5 arcmin x 5 arcmin region around Serpens South's central filament, with a phase center positioned at $\alpha(\text{J2000})=18^{\text{h}}30^{\text{m}}05.00^{\text{s}}, \delta(\text{J2000})=-02^{\circ}02^{'}30.0^{''}$ (\autoref{fig:serpsouth_fov}). During our hour-long observation, we switched from Serpens South to J1804 + 0101 every 10 minutes for complex gain calibrations, giving us a total of 45 minutes on source. At the time of our observation two antennas were not functioning properly, leaving us with a total of twenty five antennas.  

We manually flagged, calibrated and imaged our data with standard procedures using Common Astronomy Software Applications (CASA) 4.1.0. We also thoroughly inspected the data and manually flagged for obvious radio frequency interference (RFI). We used J1331+305 (3C286) as a flux and bandpass calibrator, and J1804+0101 as a gain and phase calibrator ($S_{4.75\text{GHz}}$ = 0.70 $\pm$ 0.02 Jy, $S_{7.25\text{GHz}}$ = 0.66 $\pm$ 0.02 Jy). We deconvolved the Stokes $I$ images with the Cotton-Schwab algorithm \citep{Schwab84} using the CLEAN method \citep{Hogbom74,Clark80}. We experimented with natural, robust and uniform weighting, and found the best compromise between noise level and source resolution using robust weighting \citep{Briggs95}, with the \emph{robust} parameter set to 0.5. The synthesized beam sizes and RMS values for our two images are detailed in \autoref{tab:image_pars}. We performed a primary beam correction on our images by dividing the CLEANed images by the modeled flux response of the antennas.

\capstartfalse
\begin{deluxetable}{cccc}[h!]
\tabletypesize{\footnotesize}
\tablecaption{Observation Summary}
\tablehead{
	\multicolumn{1}{c}{Frequency$^{a}$}			&
	\multicolumn{1}{c}{Beam Size$^{b}$}		&
	\multicolumn{1}{c}{Position Angle}			&
	\multicolumn{1}{c}{Image RMS}					\\
	\colhead{(GHz)}							&
	\colhead{(arcsec x arcsec)}					&
	\colhead{(degrees)}							&
	\colhead{($\mu$Jy beam$^{-1}$)}			}
\startdata
4.75	&	4.8 x 3.8&	14.8	&	11.1	\\[1ex]
7.25	&	3.1 x 2.5	&	13.4	&	8.5
\enddata
\tablenotetext{a}{4.75 and 7.25 GHz is equivalent to 6.31 and 4.14 cm respectively.}
\tablenotetext{b}{Using robust weighting in the CLEAN method}
\label{tab:image_pars}
\end{deluxetable}
\capstarttrue

We cross-referenced the positions of our radio sources with the 2MASS catalogue and ran extractions over the four IRAC bands of \emph{Spitzer}, the 24 $\mu$m MIPS band on \emph{Spitzer}, and the 70 $\mu$m PACS band on \emph{Herschel}. We use this wide range of infrared data to help determine if a radio source is associated with a YSO or if it is extragalactic. We used a 2 arcsecond maximum matching tolerance for these extractions. Apertures and adopted corrections were consistent with \citet{Gutermuth09}. The final errors on our infrared fluxes consist of the zero flux offset error used to convert magnitudes to fluxes (2\% contribution) and an assumed 10\% error on our final magnitudes, which encompasses the absolute flux calibration. We were able to produce at least one infrared source extraction on six of our radio sources, whose infrared spectral energy distributions are shown in \autoref{fig:infra_SEDs}.

\section{Methodology}
In choosing our sources, we restrict ourselves to a circular region centered on our phase center extending out to 50\% of the primary beam response, which equates to roughly 5 arcminutes in diameter. We identify our sources based on a visual inspection of the data, judging their appearance and strength in both our 7.25 and 4.75 GHz images and considering their spatial proximity to infrared associations. Similar to other radio studies, we consider a detection firm if it has a clear infrared association and has a radio flux of at least $4\sigma$ in either frequency band. In addition, we consider all objects with radio fluxes $>5\sigma$ as sources. Given our beam size relative to the size of our field of view, there is less than a 1\% chance that even one random Gaussian noise fluctuation would produce a signal in excess of $5\sigma$ in any one of our synthesized beams. This left us with a total of 18 radio sources in our field of view. We note that we detect an additional $\sim$6 sources at the 2-3$\sigma$ level that lie within 2 arcseconds of an infrared YSO candidate, however, we do not include them in our final source list. Further observations with higher sensitivity will likely be able to determine whether these sources are spurious, extragalactic or protostellar in nature.

After applying primary beam corrections, we used CASA's IMFIT function to fit 2D gaussians and derive integrated flux densities for each of our radio sources. Flux errors were calculated by adding three sources of error in quadrature: the flux error output from IMFIT, an assumed 5\% systematic absolute error on the flux calibration, and the percent uncertainty due random pointing errors in the individual antennas following the prescription outlined in \citet{Dzib14a}. Source positions were determined by using the centers of IMFIT's 2D Gaussian fits to the higher resolution 7.25 GHz image. 

It is unlikely that all of our radio sources will be associated with YSOs; there will be a level of contamination from background galaxies that emit in the radio. We can calculate the number of random background sources we would expect to find in our radio images. We use the formulation found in \citet{Shirley07} and \citet{Dzib13} who draw from radio studies done by \citet{Fomalont91}. The density of random background radio sources above a flux limit of $S\ \mu$Jy at 6 cm (4.9 GHz) is given by 
\begin{equation*}
n(>S) = 0.42\pm0.05\left(\frac{S}{30\ \mu \text{Jy}}\right)^{-1.18\pm0.19} \text{arcmin}^{-2}.
\end{equation*}
Therefore, the number density of sources with flux $S$ greater than 50 $\mu$Jy at 6 cm is 0.23 arcmin$^{-2}$. This leads to a 0.1\% chance that a background source falls inside a 4.8 x 3.9 arcsec synthesized beam centered on a radio source. It also gives us on average $\sim6\pm1.5$ background sources within our 5' x 5' region of interest above a 5$\sigma$ level. There are about 7 to 9 sources that we identify as extragalactic, which agrees with this analysis within two standard deviations at worst.

\begin{turnpage}
\capstartfalse
\begin{deluxetable*}{lcccccccccc}
\tabletypesize{\footnotesize}
\tablecaption{VLA Centimeter Continuum Sources}
\tablehead{
	\multicolumn{1}{l}{Source}								&
	\multicolumn{1}{c}{R.A.$^{a}$}								&
	\multicolumn{1}{c}{Decl.$^{a}$}							&
	\multicolumn{1}{c}{$S^{\text{int}}_{4.75 \text{GHz}}\ ^{b}$}			&
	\multicolumn{1}{c}{$S^{\text{peak}}_{4.75 \text{GHz}}\ ^{c}$}				&
	\multicolumn{1}{c}{$S^{\text{int}}_{7.25 \text{GHz}}$}				&
	\multicolumn{1}{c}{$S^{\text{peak}}_{7.25 \text{GHz}}$}				&
	\multicolumn{1}{c}{$\alpha^{\text{int}}_{\text{radio}}\ ^{d}$}			&
	\multicolumn{1}{c}{$\alpha^{\text{peak}}_{\text{radio}}\ ^{e}$}			&	
	\multicolumn{1}{c}{Class}							&				
	\multicolumn{1}{c}{Other Names}			\\ [.25ex]	
	\colhead{}										&
	\colhead{(J2000)}								&
	\colhead{(J2000)}								&
	\colhead{($\mu$Jy)}								&
	\colhead{($\mu$Jy beam$^{-1}$)}					&
	\colhead{($\mu$Jy)}								&
	\colhead{($\mu$Jy beam$^{-1}$)}					&
	\colhead{}										&
	\colhead{}										&
	\colhead{}										&
	\colhead{}
	}
\startdata
VLA\_1	&	18 30 09.68	&	-02 00 32.8	&	749 $\pm$ 46	&	718	&	844 $\pm$ 81	&	794	&	0.28 $\pm$ 0.11	&	0.24 $\pm$ 0.12	&	Extragal.	&	J183009.68-020032.7$^{i}$\\[1ex]

VLA\_2	&	18 30 04.79	&	-02 00 34.1	&	177 $\pm$ 12	&	138	&	87 $\pm$ 10	&	72	&	-1.67 $\pm$ 0.14	&	-1.55 $\pm$ 0.18	&	Extragal.	&	---\\[1ex]

VLA\_3	&	18 30 02.55	&	-02 00 48.5	&	140 $\pm$ 8	&	99	&	116 $\pm$ 14	&	87	&	-0.46 $\pm$ 0.14	&	-0.3 $\pm$ 0.18	&	Extragal.	&	---\\[1ex]

VLA\_4	&	18 30 01.53	&	-02 00 51.6	&	1663 $\pm$ 101	&	1178	&	946 $\pm$ 84	&	515	&	-1.33 $\pm$ 0.11	&	-1.96 $\pm$ 0.1	&	Extragal.	&	---\\[1ex]

VLA\_5	&	18 30 14.79	&	-02 01 27.3	&	106 $\pm$ 9	&	95	&	167 $\pm$ 20	&	173	&	1.07 $\pm$ 0.15	&	1.41 $\pm$ 0.18	&	Extragal.?	&	---\\[1ex]

VLA\_6	&	18 30 06.62	&	-02 01 28.9	&	79 $\pm$ 8	&	61	&	33 $\pm$ 5	&	56	&	-2.04 $\pm$ 0.18	&	-0.22 $\pm$ 0.25	&	Extragal.	&	---\\[1ex]

VLA\_7	&	18 30 05.81	&	-02 01 44.9	&	48 $\pm$ 4	&	40	&	84 $\pm$ 9	&	63	&	1.29 $\pm$ 0.15	&	1.08 $\pm$ 0.32	&	Class II	&	---\\[1ex]

VLA\_8	&	18 30 06.89	&	-02 02 11.5	&	39 $\pm$ 4	&	46	&	50 $\pm$ 6	&	50	&	0.55 $\pm$ 0.17	&	0.18 $\pm$ 0.3	&	Extragal.?	&	---\\[1ex]

VLA\_9	&	18 30 02.80	&	-02 02 25.3	&	66 $\pm$ 3	&	50	&	60 $\pm$ 3	&	44	&	-0.22 $\pm$ 0.08	&	-0.31 $\pm$ 0.3	&	Unknown	&	---\\[1ex]

VLA\_10	&	18 30 02.95	&	-02 02 32.2	&	287 $\pm$ 16	&	119	&	260 $\pm$ 22	&	73	&	-0.23 $\pm$ 0.1	&	-1.15 $\pm$ 0.17	&	Unknown	&	---\\[1ex]

VLA\_11	&	18 30 03.36	&	-02 02 45.8	&	38 $\pm$ 3	&	36	&	73 $\pm$ 5	&	58	&	1.5 $\pm$ 0.11	&	1.14 $\pm$ 0.35	&	Class I	&	P2$^{g}$,CARMA-5$^{h}$\\[1ex]

VLA\_12	&	18 30 04.11	&	-02 03 02.1	&	30 $\pm$ 3	&	24	&	79 $\pm$ 4	&	58	&	2.31 $\pm$ 0.12	&	2.06 $\pm$ 0.49	&	Class 0	&	MM18$^{f}$,CARMA-7$^{h}$\\[1ex]

VLA\_13	&	18 30 03.54	&	-02 03 08.4	&	186 $\pm$ 9	&	178	&	231 $\pm$ 13	&	227	&	0.51 $\pm$ 0.08	&	0.58 $\pm$ 0.11	&	Class 0/I	&	P3$^{g}$,CARMA-6$^{h}$\\[1ex]

VLA\_14	&	18 30 03.84	&	-02 03 12.2	&	31 $\pm$ 1	&	37	&	83 $\pm$ 5	&	61	&	2.29 $\pm$ 0.09	&	1.19 $\pm$ 0.34	&	Class 0	&	CARMA-6$^{h}$\\[1ex]

VLA\_15	&	18 30 04.60	&	-02 03 14.1	&	68 $\pm$ 3	&	52	&	49 $\pm$ 4	&	51	&	-0.74 $\pm$ 0.11	&	-0.08 $\pm$ 0.28	&	Extragal.	&	---\\[1ex]

VLA\_16	&	18 30 03.25	&	-02 03 26.6	&	63 $\pm$ 3	&	58	&	55 $\pm$ 3	&	65	&	-0.28 $\pm$ 0.09	&	0.27 $\pm$ 0.24	&	Class I	&	---\\[1ex]

VLA\_17	&	18 30 01.32	&	-02 03 42.9	&	146 $\pm$ 8	&	135	&	161 $\pm$ 12	&	167	&	0.22 $\pm$ 0.1	&	0.5 $\pm$ 0.13	&	Flat/II	&	Y1$^{g}$\\[1ex]

VLA\_18	&	18 30 09.40	&	-02 04 08.8	&	170 $\pm$ 10	&	182	&	167 $\pm$ 18	&	134	&	-0.04 $\pm$ 0.12	&	-0.71 $\pm$ 0.14	&	Extragal.?	&	---

\enddata
\tablenotetext{a}{Centers of 2D Gaussian fits for sources in 7.25 GHz map using CASA's IMFIT procedure.}
\tablenotetext{b}{Integrated flux values and errors using IMFIT from images deconvolved with Briggs weighting, \emph{robust}=0.5 \citep{Briggs95}.}
\tablenotetext{c}{Peak flux values calculated by taking the brightest pixel at the source's center.}
\tablenotetext{d}{Spectral index of integrated flux from 7.25 to 4.75 GHz, see \autoref{sec:spix} for details on calculation.}
\tablenotetext{e}{Spectral index of peak flux from 7.25 to 4.75 GHz.}
\tablenotetext{f}{From \citet{Maury11}}
\tablenotetext{g}{From \citet{Teixeira12}}
\tablenotetext{h}{From \citet{Plunkett15}}
\tablenotetext{i}{From \citet{OrtizLeon15}}
\label{tab:flux_table}
\end{deluxetable*}
\capstarttrue
\end{turnpage}

\subsection{Radio and Infrared Spectral Indices}
\label{sec:spix}

Radio emission from a YSO can have two components: a thermal component coming from free-free bremsstrahlung of an ionized region and a non-thermal component generally in the form of gyrosynchrotron emission created by magnetic fields in the stellar corona. While high mass stars have strong enough internal luminosities to support compact HII regions, low mass stars generally do not. Thermal emission from ionized regions around low-mass YSOs is typically thought to be driven by collimated outflows that shock the surrounding material \citep{Anglada98}. Because outflows are common among less-developed YSOs, thermal radio emission is found predominately around the earliest protostellar phases, Class 0 and I. Non-thermal emission is generally found around more developed pre-main sequence T Tauri stars (Class II \& III) because their stellar coronae are exposed. There are a few cases, however, where non-thermal emission has been found around young protostars, which could in part be due to geometric effects or envelope-clearing by a companion star \citep{Dzib10}. This suggests that young YSOs may in fact emit copious amounts of non-thermal emission that is then absorbed by thermal emission coming from larger size scales around the protostar.

At radio wavelengths, the spectral index is defined as the difference of the natural log of the radio flux at two different radio wavelengths, divided by the difference of the natural log of those wavelengths:
\begin{equation}
\label{eq:radio_spix}
\alpha_{\text{radio}} = \frac{\ln(S_{\lambda_{1}}/S_{\lambda_{2}})}{\ln({\lambda_{2}/\lambda_{1}})},
\end{equation}
where $\lambda_{2} > \lambda_{1}$ \citep{Shirley07}. Note that $S_{\lambda}$ is defined as a flux density and not a flux (e.g. $\lambda\cdot S_{\lambda}$ or $\nu\cdot S_{\nu}$). 

Flat radio spectral indices ($\alpha > -0.1$) are generally indicative of optically thin thermal free-free emission from an ionized plasma, while rising indices ($\alpha \sim 2.0$) are representative of the optically thick case \citep{Ghavamian98}. Steeply falling radio spectral indices ($\alpha \sim -2.0$) are common from sources of non-thermal radio emission (e.g. AGN). Because of this, radio spectral indices are commonly used to help discriminate between thermal and non-thermal emission processes; however, this can be complicated due to errors on measured fluxes and the blending of multiple emission components. Other ways to deduce the emission mechanism of a radio source is to measure the variability of its radio flux over time and to probe for polarization, neither of which we can do in this present study.

\begin{figure*}
\label{fig:infra_SEDs}
\centering
\includegraphics[keepaspectratio=True,scale=.55]{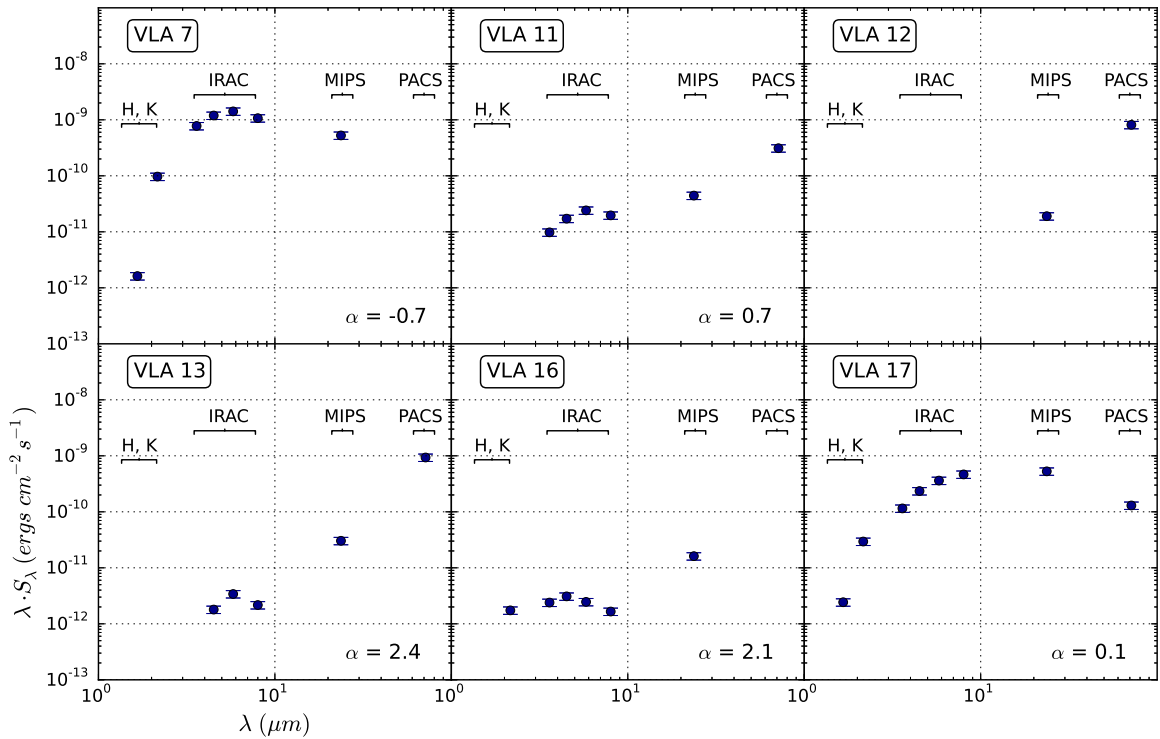}
\caption{\small{SEDs of all protostellar candidates that have infrared associations. Infrared bands are 2MASS H and K from 1.6 - 2.2 $\mu$m, IRAC 1, 2, 3 and 4 from 3.6 - 8.0 $\mu$m, MIPS 24$\mu$m and PACS 70$\mu$m. Our derived infrared spectral index from 8 to 24 $\mu$m is shown for sources with successful fits in those bands and have errors of $\pm0.2$.
}}
\end{figure*}

In the infrared, four main categories of YSOs exist based on the shape of their spectrum: Class I, Flat-Spectrum, Class II and Class III YSOs. These classifications roughly correspond to unique development phases in chronological order; however, it is good to note that a direct connection between infrared class and unique evolutionary stage can be confused by possible misclassifications based on inclination effects \citep{Crapsi08}. \citet{Greene94} presented YSO classifications based on infrared spectral indices,
where the infrared spectral index $\alpha_{\text{IR}}$ is defined as
\begin{equation}
\label{infra_spix}
\alpha_{\text{IR}} = \frac{\ln\left((\lambda_{2}\cdot S_{\lambda_{2}})/(\lambda_{1}\cdot S_{\lambda_{1}})\right)}{\ln({\lambda_{2}/\lambda_{1}})}:
\end{equation}
\begin{align}
&\text{Class I}  & 0.3\le&\alpha_{\text{IR}}\\
&\text{Flat Spectrum} & -0.3\le&\alpha_{\text{IR}}<0.3\\
&\text{Class II} & -1.6\le&\alpha_{\text{IR}}<-0.3\\
&\text{Class III}  & &\alpha_{\text{IR}}<-1.6,
\end{align}
with $\lambda_{2} > \lambda_{1}$. We derive our infrared spectral indices between the IRAC 8 $\mu$m band and the MIPS 24 $\mu$m band. Our infrared spectral indices have a standard error of $\pm0.2$ based on an error analysis that combines in quadrature the zero flux offset error used to convert magnitudes to fluxes (2\% contribution) and an assumed 10\% error in our extracted magnitudes. There are five sources where we have successful source extractions at 8 and 24 $\mu$m. \autoref{fig:infra_SEDs} shows these SEDs and our derived infrared spectral indices. 

\section{Results: Classifications and Radio Properties}
\label{sec:results}
\newcommand{\sourcenote}[1]{\textit{#1 ---}}

We detail the properties of our eighteen radio sources in \autoref{tab:flux_table}, which includes their integrated and peak radio fluxes, derived spectral indices and classifications based on our radio and infrared data. Other detections of our sources in previous works are also discussed here, which come primarily from four studies: a near-infrared spectral line study of molecular hydrogen jets and outflows done by \citet{Teixeira12} (referred to as \citetalias{Teixeira12}), a 1.2 millimeter continuum survey done by \citet{Maury11} (referred to as \citetalias{Maury11}), a 2.7 millimeter spectral line study of molecular outflows done by \citet{Plunkett15} (referred to as \citetalias{Plunkett15}), and a centimeter radio continuum study with the VLA done by \citet{OrtizLeon15} (referred to as \citetalias{OrtizLeon15}). The major differences between our study and the VLA study of \citetalias{OrtizLeon15} are (1) they had higher angular resolution by over a factor of ten, and (2) they had lower point source sensitivity by over a factor of two. Their lower sensitivity did not enable them to make any firm radio detections of any YSO candidates in our field of view. For radio sources we report as having radio fluxes within their detectable limits, their non-detection could be due to either their small beam size over-resolving the emission structure, time variability in the source's emission, or both. For the case of the former, this would suggest that there is a significant source of thermal emission originating at or above size scales of hundreds of AU: the physical distance corresponding to the angular separation of \citetalias{OrtizLeon15}'s 0.3 arcsecond synthesized beam assuming a distance of 429 pc to the cloud.

\subsection{Protostellar Sources}
\begin{figure*}
\label{fig:infrared_associations}
\centering
\includegraphics[keepaspectratio=True,scale=.65]{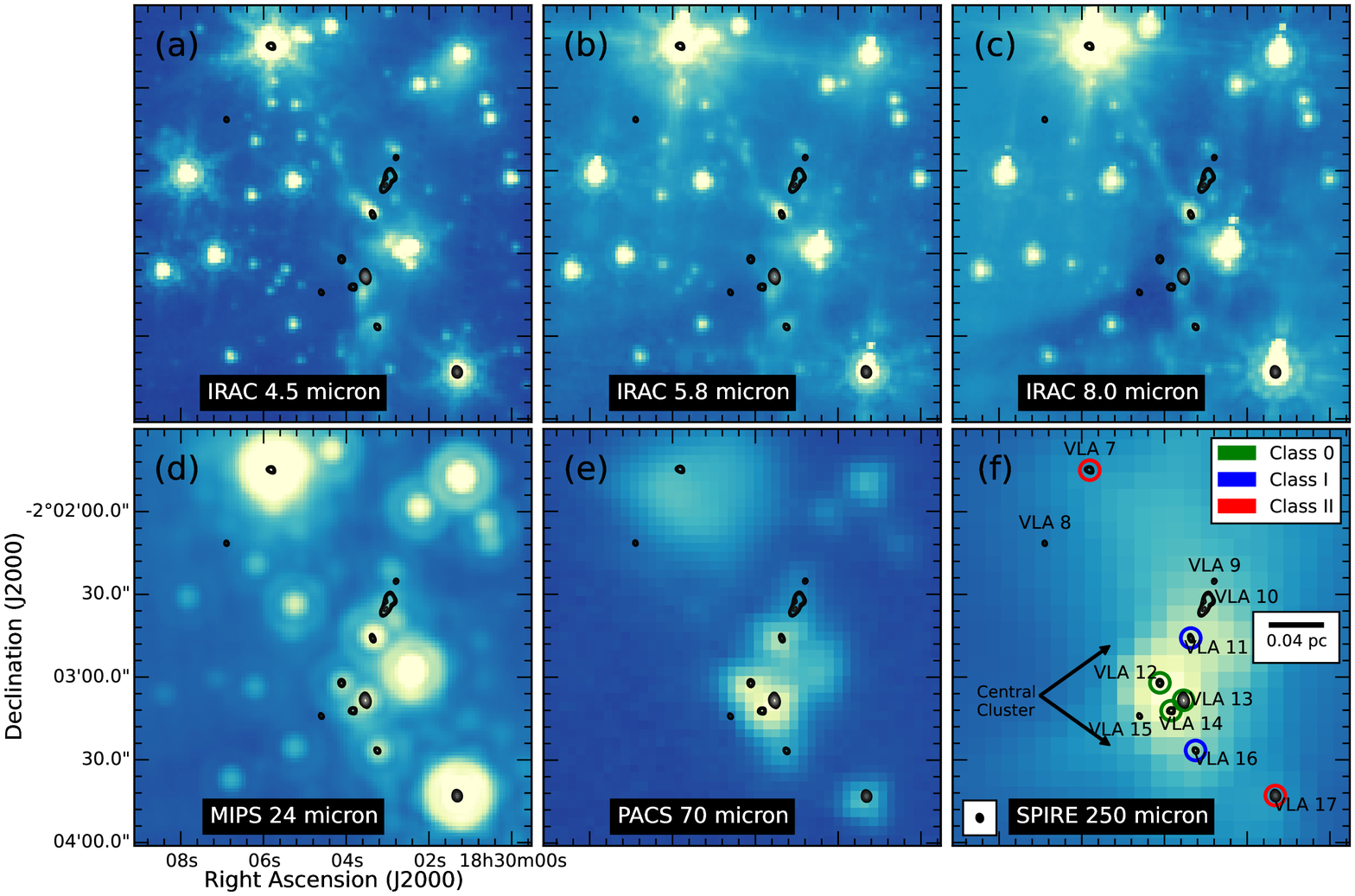}
\caption{\small{7.25 GHz (4.14 cm) radio contours plotted over infrared and far-infrared images ranging from 4.5 to 250 $\mu$m. Radio contours start at the 4$\sigma$ level, with the deconvolved beam size shown at lower-left. {\bfseries (a)} \emph{Spitzer} IRAC 4.5 $\mu$m image. {\bfseries (b)} \emph{Spitzer} IRAC 5.8 $\mu$m image. {\bfseries (c)} \emph{Spitzer} IRAC 8.0 $\mu$m image. {\bfseries (d)} \emph{Spitzer} MIPS 24 $\mu$m image. {\bfseries (e)} \emph{Herschel} PACS 70 $\mu$m image. {\bfseries (f)} \emph{Herschel} SPIRE 250 $\mu$m image. Note how the radio sources in the central cluster are not visible in the near infrared images, but start to become visible long-ward of 24 microns.
}}
\end{figure*}

\begin{figure*}
\label{fig:vla10 and vla12}
\epsscale{1.17}
\plottwo{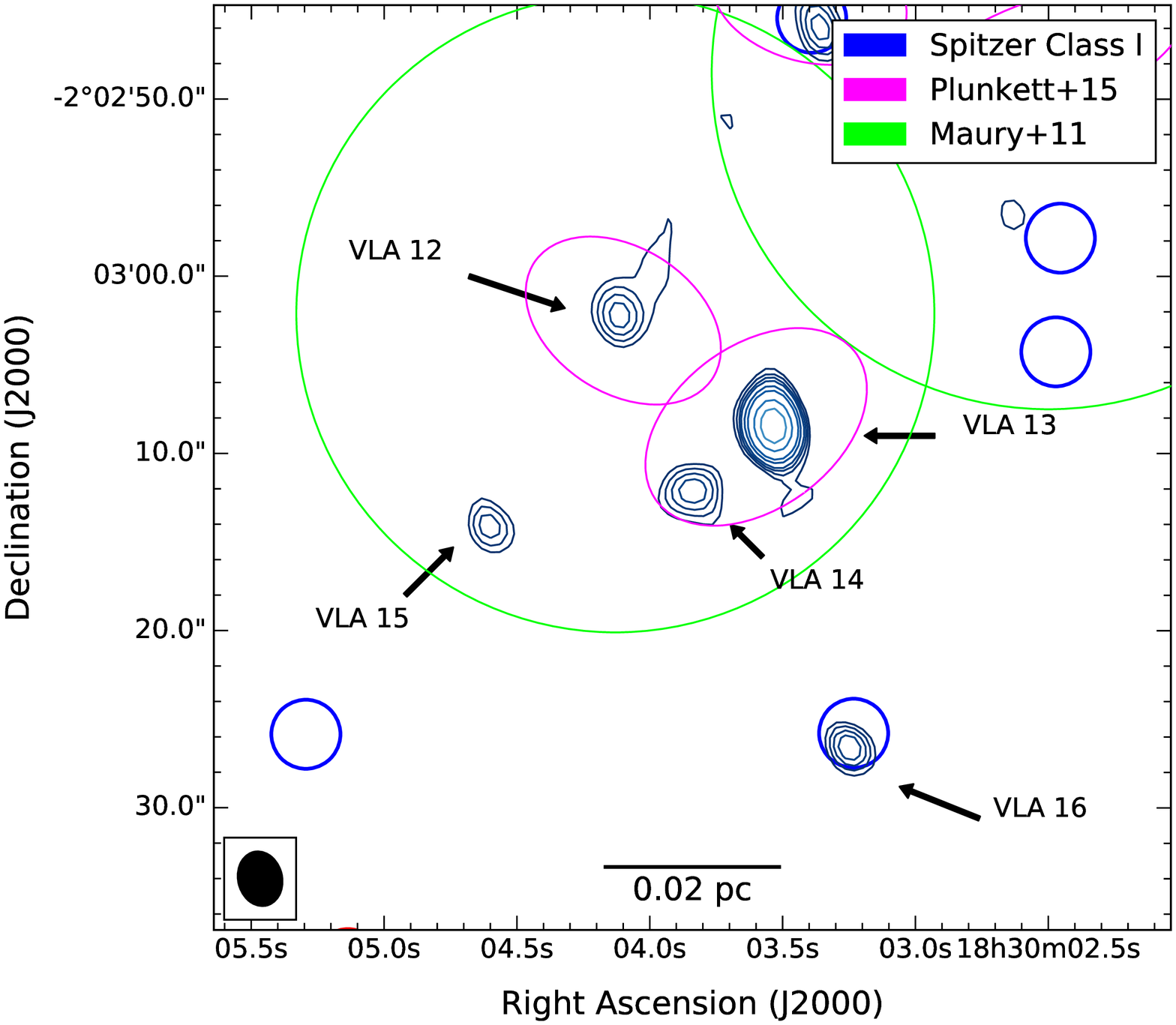}{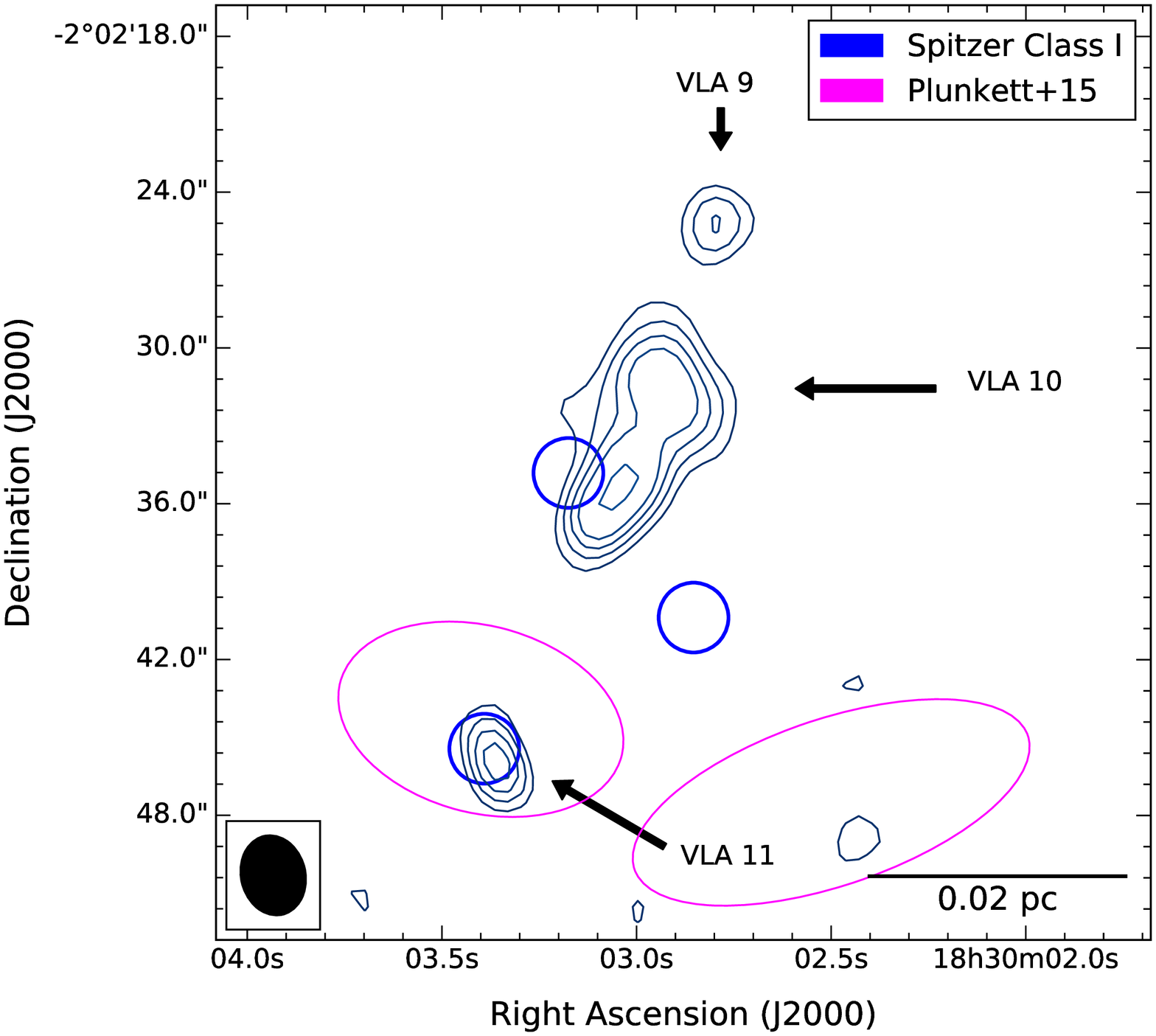}
\caption{
{\bfseries Left}: Zoom-in of VLA 12 and surrounding sources. {\bfseries Right:} Zoom-in of the elongated VLA 10 source. In both images, the blue solid contours represent the 7.25 GHz radio signal and start at the 3$\sigma$ level and increase to 4, 5, 6, 8, 10, 15, 20, 30, and 50$\sigma$. The blue circles are \emph{Spitzer} identified Class I sources (Gutermuth et al. in prep.). Green circles are the FWHM fits of the 1.2 mm emission peaks from \citet{Maury11}. Magenta circles are fits to the 2.7 mm emission from \citet{Plunkett15}. Projected size scale assumes a distance of 429 pc to Serpens South.
}
\end{figure*}

Seven of our eighteen radio sources, VLA 7, 11, 12, 13, 14, 16 and 17, are likely to be protostellar in nature. All except VLA 14 lie within 2 arcseconds of an infrared/submillimeter-identified Class 0, I or II YSO. All sources have either flat or rising radio spectral indices indicative of optically thin or thick thermal emission expected from a young YSO. \autoref{fig:infrared_associations} shows our 7.25 GHz radio contours against various infrared and submillimeter images of the core of Serpens South. \autoref{fig:vla10 and vla12} shows a zoom-in on our 7.25 GHz and 4.75 GHz radio contours for most of these sources along with \emph{Spitzer} Class I \& II YSOs (Gutermuth et al. in prep). 

\emph{VLA 7} is spatially associated with a Class II YSO in the infrared that has a calculated $\alpha_{\text{IR}}$ of -0.7. It has a radio spectral index of $\sim$1.3 indicative of a mix of optically thin and thick thermal emission. 

\emph{VLA 11} is associated with an infrared source with $\alpha_{\text{IR}}$=0.7, which makes it a Class I protostar. In the IRAC bands, VLA 11's SED seems to be turning over, but its MIPS and PACS emission clearly shows it has a rising SED at mid-infrared wavelengths. 

VLA 11 was also found to be driving a jet. \citetalias{Teixeira12} found a molecular hydrogen emission-line object (MHO) feature that they conclude is driven by the infrared source associated with our VLA 11 source, which they call ``P2.'' We note that while \citetalias{Teixeira12} claims this YSO to be Class 0 by citing the \emph{Herschel} study of \citet{Bontemps10}, there was no officially published catalogue of \emph{Herschel} protostars from Serpens South at the time. Currently, only a catalogue of dense cores and YSO candidates without classifications exists for \emph{Herschel} data in the Aquila region \citep{Konyves15}.

\citetalias{Maury11} images a millimeter peak, called ``MM16,'' centered East of VLA 11 but encompasses VLA 11 within its FWHM fit. While they suggest MM16 is a blending of multiple sources, we definitively show that there are indeed at least two sources associated with the their millimeter peak. \citetalias{Plunkett15} also showed this with higher resolution mm continuum imaging, detecting both VLA 11, which they call ``CARMA-5'', and the source at the center of \citetalias{Maury11}'s MM16.

\emph{VLA 12} is not detected short-ward of 24 microns, meaning it is a highly embedded source. It is, however, the peak structure of diffuse 1.2 mm emission and was therefore classified as a Class 0 source \citep{Maury11}. This classification used a 25 arcsecond FWHM fit, which also encompassed the infrared sources associated with VLA 13 and 14 (see \autoref{fig:vla10 and vla12}). Here, we reaffirm VLA 12's Class 0 classification with the detection of thermal radio emission indicative of outflows and the presence of a central YSO \citep{Andre00}. Recent ALMA data also finds VLA 12 to be a driver of a highly collimated, episodic molecular outflow \citep{Plunkett15b}, running 4$^{\circ}$ East of the North-South vertical direction. VLA 12 is also detected by \citetalias{Plunkett15} at 2.7 mm, who resolve VLA 12's emission from its nearby companions VLA 13 and VLA 14 (referred to as CARMA-7). At the 3$\sigma$ level, VLA 12's radio emission has elongated structure towards the North-West (\autoref{fig:vla10 and vla12}), which aligns with its continuum emission at 1 mm \citep{Plunkett15b}.

\emph{VLA 13} is associated with an infrared source that has a full set of infrared extractions except for the IRAC 1 band at 3.6 $\mu$m (\autoref{fig:infra_SEDs}). The lack of a flux at the IRAC 1 band, as noted by \citetalias{Plunkett15}, would prevent it from being classified as a YSO candidate by \citet{Evans09}. However, because we (1) have fluxes associated with VLA 13 from the IRAC 2 band to the PACS 70$\mu$m band and (2) have detected thermal radio emission, we conclude that VLA 13 is likely an embedded Class 0/I protostar. It shows comparatively strong radio emission suggesting it could be driving a relatively powerful outflow, perhaps being the most dominant driver of outflows in the densely populated region of Serpens South's core. 

\emph{VLA 14} is the only YSO candidate we detect that has yet to be definitively detected by any previous study. In far-infrared images from 24 to 250 $\mu$m it is clear that, in projection, VLA 14 lies within a gaseous envelope (\autoref{fig:infrared_associations}). However, because of the low resolution of the far infrared images that blends the emission of VLA 13 with VLA 14, we do not retrieve any source extraction within 2 arcseconds of VLA 14's radio peak. Even in the higher resolution MIPS 24 $\mu$m image, VLA 14 is not seen as an individual source but rather an extension of the emission structure from VLA 13 (\autoref{fig:infrared_associations}). This is also seen in \citetalias{Plunkett15}'s mm continuum emission structure. In the near-infrared, VLA 14 is too deeply embedded to be seen. This is likely why previous studies missed this source and it implies that if a central source exists it is likely young and/or a low luminosity object. There are three main reasons why we conclude that VLA 14 is very likely an embedded Class 0 protostar: (1) it has a steeply rising radio spectral index indicative of optically thick thermal radio emission, which is also suggestive of outflows and therefore a central source, (2) it is a spatially resolved structure from VLA 13 in both our 7.25 and 4.75 GHz radio images, and (3) it is set against strong and extended submillimeter emission indicative of an extended envelope: the same millimeter emission from \citet{Maury11} (SerpS-MM18) that blends VLA 14 with both VLA 12 and 13 and the submillimeter emission from \citet{Plunkett15} (CARMA-6) that blends VLA 14 with VLA 13.

\emph{VLA 16} is associated with an infrared source that has a steeply rising infrared spectral index and is therefore classified as a Class I YSO, although just looking at its SED at IRAC wavelengths it may be confused as a Flat-Spectrum or Class II source. It has a relatively flat radio spectral index suggesting optically thin thermal radio emission.

\emph{VLA 17}'s infrared SED suggests it is a Flat-Spectrum source that may be transitioning to a Class II source, due to the fact that its 70 $\mu$m flux is significantly lower than its 24 $\mu$m flux. It has comparatively strong radio flux with a slightly rising radio spectral index, which, similar to VLA 13, suggests that it could be a dominant contributor to large-scale outflow from the core. VLA 17 is also detected by the CARMA study, which they label ``CARMA-1,'' and is also claimed to be a driver of a MHO feature referred to as ``Y1'' by \citet{Teixeira12}.

\subsection{Extragalactic Sources}
\begin{figure}[h!]
\label{fig:vla4}
\centering
\includegraphics[keepaspectratio=True,scale=.325]{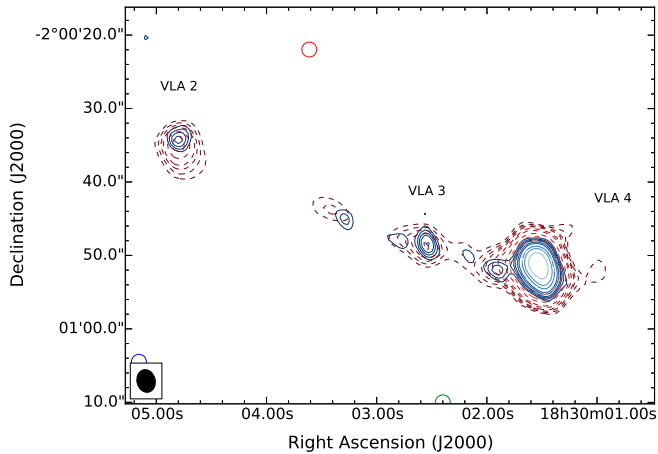}
\caption{\small{ Zoom-in on sources VLA 2, 3 and 4. Blue solid and red dashed contours correspond to 7.25 and 4.75 GHz radio images respectively, which start at their respective 3$\sigma$ levels. This system of extragalactic sources could be in fact one central galaxy with powerful radio lobes.
}}
\end{figure}

VLA 2, 3, 4, 6, and 15 are very likely extragalactic in nature. They have no visible associations from 1 $\mu$m to 70$\mu$m and have negative radio spectral indices indicative of non-thermal synchrotron emission typical of a galaxy's radio spectrum. VLA 2, 3 and 4 are interestingly aligned along an axis (\autoref{fig:vla4}). 
One possible explanation is that these three sources actually form one source: one side of a radio jet emanating from the central galaxy (VLA 2) and ending in a radio lobe (VLA 4). These strong sources of radio emission were not detected by the high resolution radio survey done by \citetalias{OrtizLeon15}. The most probable reason why \citetalias{OrtizLeon15} did not detect these strong sources is because they are extended sources, meaning that the high resolution ($\sim$0.3$^{"}$) synthesized beam of \citetalias{OrtizLeon15} resolved out the radio flux to a level below their point source sensitivity.

While VLA 1 and 5 have no infrared, far-infrared or millimeter associations and are isolated from the Serpens South filament, they have flat and rising radio spectral indices suggesting optically thin or thick free-free emission respectively. Although galaxies can produce free-free emission from large HII regions, it is generally a sub-dominant component compared to its synchrotron output when an AGN is present. Moreover, in \citet{Dzib13}'s study of 190 radio sources in the Ophiuchus star forming cloud, they did not find a single extragalactic source with a radio spectral index larger than $\sim$0. However, we should note that VLA 1 and 5 are relatively far from our phase center ($\theta\sim 2.3^{'}$) meaning that VLA pointing errors could be larger than accounted for in our error analysis. We tentatively classify these sources as extragalactic due to their lack of any association at shorter wavelengths.  

\subsection{Unclear Sources}
We are unsure as to the nature of VLA 10 (\autoref{fig:vla10 and vla12}). We treat the whole elongated structure as one source. At our higher resolution 7.25 GHz image, we can resolve two distinct emission components that are blended at our lower resolution 6.3 cm image. Although our measured fluxes are uncertain due to source blending at 4.75 GHz, VLA 10's peak flux derived spectral index is strongly negative suggesting non-thermal emission and an extragalactic classification. However, VLA 10's Southern structure is also within 2 arcseconds of a \emph{Spitzer} identified Class I source. This raises the possibility that the Northern structure is a background source while the Southern structure is a YSO radio source. Another possibility is that VLA 10 is representative of post-shock free-free emission from an outflow emanating from either the Southern VLA 10 structure, the Northern VLA 10 structure, or from sources VLA 11, VLA 9, or another source. Indeed, recent millimeter studies tracing $^{12}$CO outflow emission shows a low-velocity redshifted outflow structure that aligns spatially and morphologically with VLA 10's radio emission \citep{Plunkett15}.

Further evidence that would support a non-extragalactic classification of both the South and North structures of VLA 10 is the fact that the dust emission seen at and longward of 70 $\mu$m neatly fits around VLA 10's elongated morphology, as seen in \autoref{fig:infrared_associations}-(e). Lastly, its elongation towards the North-West direction aligns with the direction of the large-scale red-shifted molecular outflow seen in \citetalias{Plunkett15}.
Higher resolution, multi-epoch observations should be able to spatially separate the individual emission components and better determine the nature behind their radio emission.

Similar to VLA 10, the nature of VLA 9 is unclear. Although it has no strong infrared associations, it does lie within 5 arcseconds of a weak infrared source. It is placed in a region of high YSO surface density, falls in front of diffuse, far-infrared emission and has a relatively flat spectral index: increasing the chances it is a YSO candidate. Its 4.75 GHz emission blends with VLA 10 at the 3$\sigma$ level. While this could be due to blending of VLA 9 and 10's emission structure due to the larger synthesized beam at 4.75 GHz, it does raise the possibility that VLA 9 is tied to the structure powering VLA 10's radio emission (\autoref{fig:vla10 and vla12}).

VLA 18 is spatially distinct from the filament and has a flat radio spectral index of about $0\pm0.1$, consistent with optically thin thermal radio emission driven by protostellar outflows. It has three extremely weak infrared associations at the IRAC infrared bands from 3.6 to 8.0 $\mu$m, the 24 $\mu$m MIPS band, and a weak and diffuse association at 100 and 160 $\mu$m from \emph{Herschel} images. However, while these associations are visible upon manual inspection of the data, the only ones strong and compact enough to enable successful source extractions are the IRAC band associations. VLA 18 also lies $\sim$3 arcseconds from a \emph{Herschel}-identified starless core \citep[see][]{Konyves15}, which further supports a YSO classification. Currently, we do not have enough evidence to firmly support a protostellar classification due to the lack of far infrared associations, however, higher resolution millimeter and radio studies could clarify the nature of this source.

\section{Discussion}
\label{sec:discussion}

The Serpens South region is filled with YSOs identified in the infrared. \emph{Spitzer} images reveal upwards of 91 YSOs in the entire Serpens South cloud, 59\% of which are Class I protostars \citep{Gutermuth08}. At the core, this fraction rises to 77\%, suggesting a recent onset of star formation in this concentrated area (within about $2\times10^{5}$ years). \citet{Gutermuth08} define the core of Serpens South within a circular region centered at $\alpha$=18.30.03 and $\delta$=-02.01.58.2 and extending out by 2.5 arcminutes, which is similar in both position and size to our radio observation. Within this spatially defined core lies a strong peak in sub-millimeter emission, corresponding to the highly clustered radio sources of VLA 11 to VLA 16 (see \autoref{fig:infrared_associations}-f).

The three Class 0 sources in the central cluster lie within a projected separation of only 10 arcseconds, which corresponds to 0.04 pc or roughly 4,100 AU, although they may lie further away in 3D coordinates. The fact that we detect three clustered Class 0 protostars suggests that the onset of star formation in the central cluster is very recent, on timescales of the Class 0 phase duration of $\sim10^{5}$ years \citep{Dunham15}.

\begin{figure*}
\label{fig:radio_corr}
\centering
\epsscale{1.15}
\plottwo{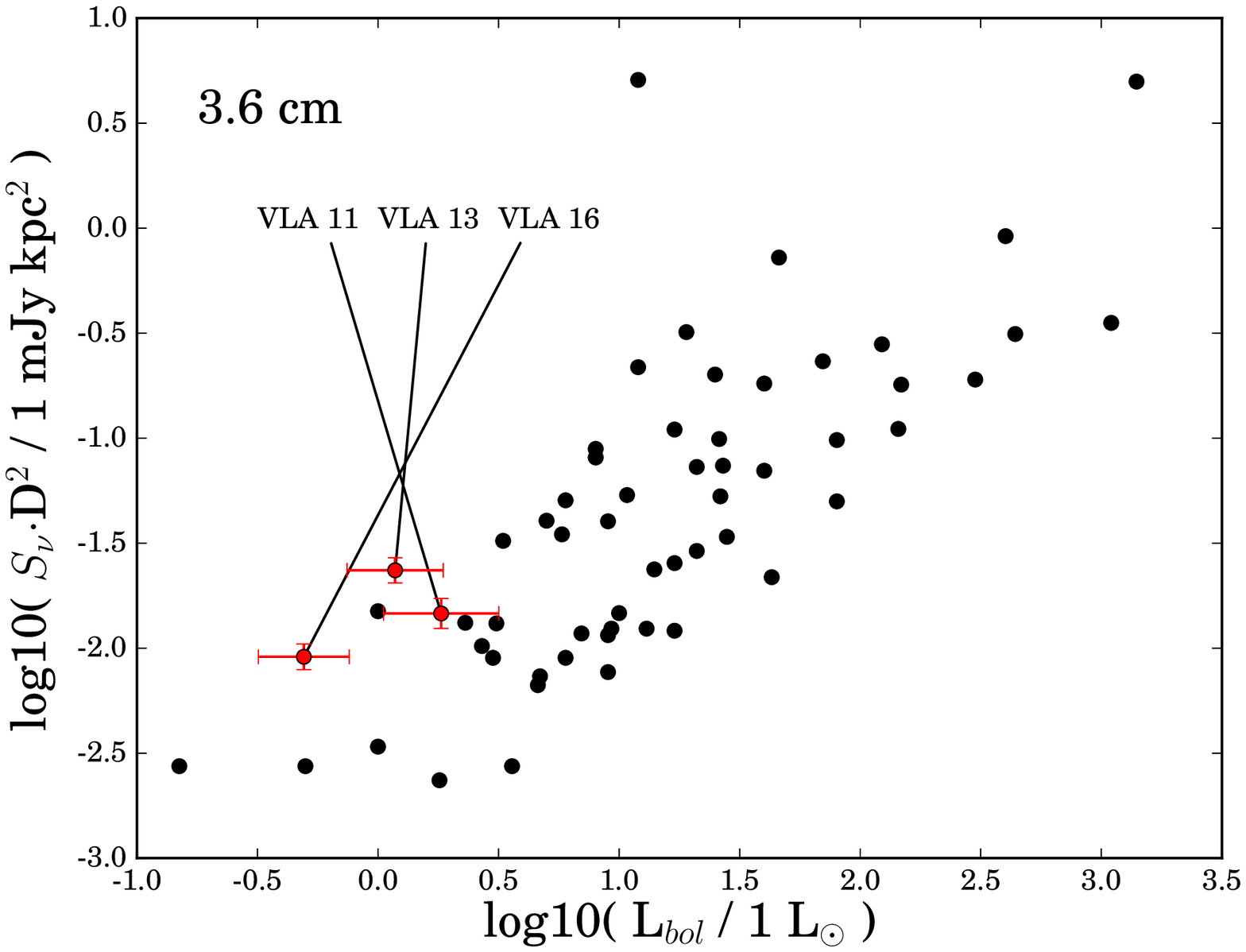}{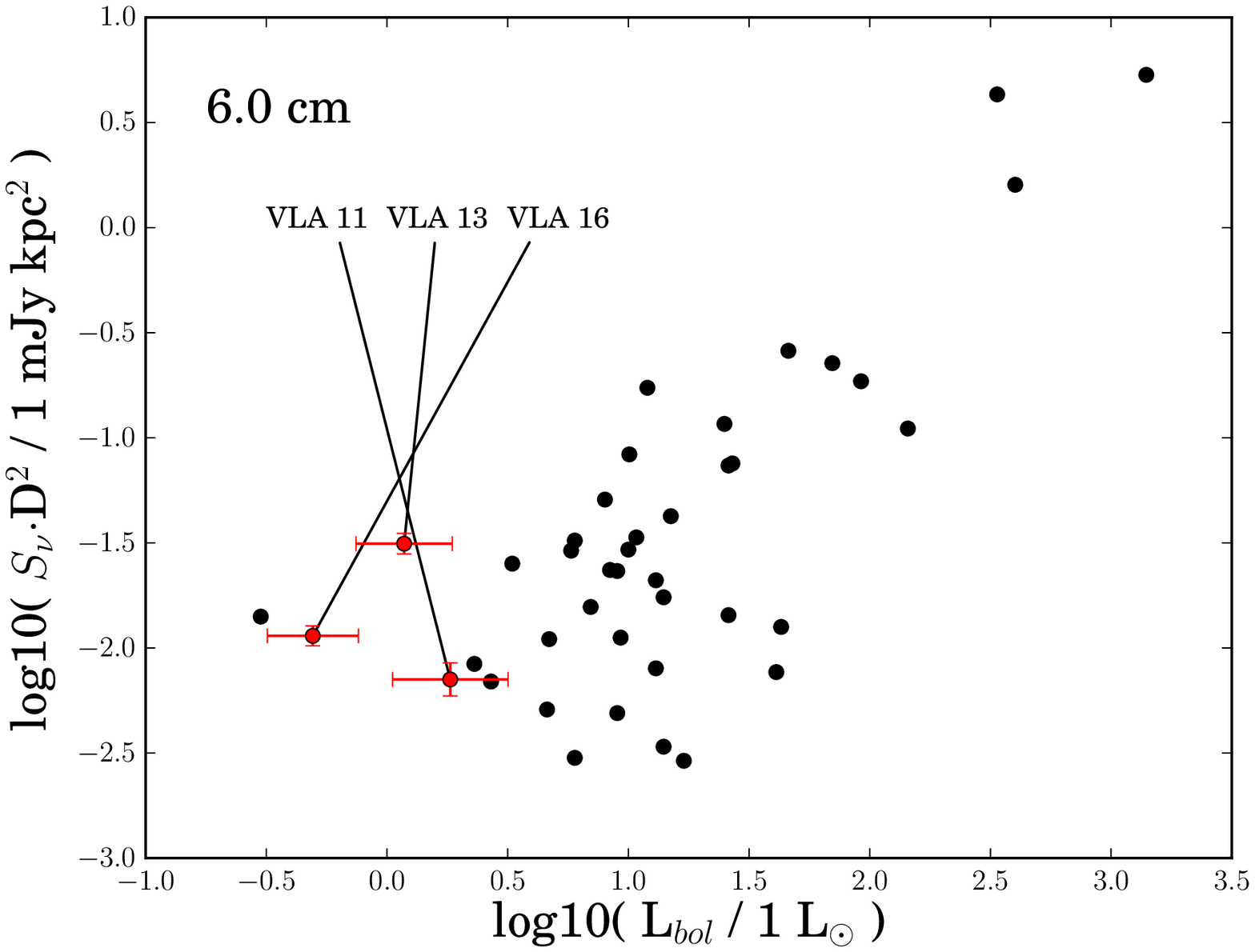}
\caption{\small{Figure 2 from \citet{Shirley07} showing the 3.6 cm vs. bolometric luminosity correlation ({\bfseries left}) and 6.0 cm vs. bolometric luminosity ({\bfseries right}) for protostars. Three of our protostellar sources are plotted with bolometric luminosities derived from a correlation that utilizes the protostars' mid-infrared luminosity \citep[see][]{Kryukova12,Dunham13}. Figure reproduced from \citet{Shirley07} with permission.
}}
\end{figure*}

Out of the roughly 35 Class I protostars in our field of view, we detect radio flux from only six of them. In addition, the radio fluxes of our YSO sources are for the most part weak, on the order of or less than one hundred micro-Janksy. 
\citet{OrtizLeon15} suggested that, assuming a spherical wind model and using \citet{Rodriguez89} as a guide, young YSOs in Serpens South are expected to emit about 0.34 mJy of thermal radio flux given a distance of 429 pc. If that distance is increased to 700 pc, that expectation falls to 0.12 mJy, which is out of their detection range but still well within ours, possibly explaining their lack of any YSO detections in the core of Serpens South. The fact that we only pick up a handful of detections with over a factor of two increase in sensitivity over \citet{OrtizLeon15} suggests that the distance to Serpens South is likely not what is causing our lack of YSO radio detections (assuming a spherical wind model), otherwise we would have likely detected many more Class I sources in the radio.

At high resolution scales, the collimated outflows that are the energy sources of thermal radio emission can be resolved \citep[e.g.][]{Rodriguez03}. Although our spatial resolution is not high enough to probe the inner regions of outflows, even at 3 arcsecond resolution scales we recover some source morphology for VLA 11 in our 7.25 GHz map; two-component Gaussian fitting with the IMFIT task on our 7.25 GHz image returns an estimated deconvolved beam size of 3.1$\pm$0.2 arcsec $\times$ 0.8$\pm$0.3 arcsec with a position angle of 11$\pm$2 degrees for VLA 11. This morphology also is in the same direction of MHO features that \citet{Teixeira12} concluded were being driven by VLA 11. Further high resolution outflow studies may be able to clarify this connection.

Because thermal radio emission is driven by outflows, the strength of YSOs' relative radio flux can be used as a rough proxy for relative outflow strength. VLA 13 has the strongest radio flux out of our YSO candidates and it is centrally located within the core of Serpens South, where there is strong $^{12}$CO emission tracing molecular outflows. We therefore speculate that it could be driving a relatively strong outflow and could be a strong contributor to Serpens South's large scale molecular outflow seen in the millimeter regime \citep[see][]{Nakamura11,Plunkett15}.

\subsection{The $S_{\text{radio}}$ and $L_{\text{bol}}$ Relationship}
\label{sec:correlations}
A correlation between centimeter flux and bolometric luminosity is known to exist for low and intermediate mass protostars. This correlation is related to another relationship, the outflow force versus bolometric luminosity relationship, which shows that higher luminosity YSOs drive more powerful outflows \citep{Wu04}. Because thermal radio emission is driven by outflows and stronger outflows produce more thermal emission \citep{Ghavamian98}, the centimeter flux versus bolometric luminosity relation is naturally an extension of the former relationship. \citet{Anglada95} was one of the first to relate a YSO's centimeter continuum emission to its bolometric luminosity. Since then, this relationship has been updated and improved upon with protostars at lower and lower bolometric luminosities \citep[e.g.,][]{Anglada98,Beltran01,Shirley07,Scaife11a}. The intrinsic scatter of this relationship is likely affected by emission variability.

The bolometric luminosity of a protostar is dominated by emission from its envelope at sub-millimeter wavelengths. However, in the core of Serpens South, it is not possible to construct sub-millimeter SEDs for individual protostars because they are blended in the \emph{Herschel} images. To get an estimate of their bolometric luminosities, we have relied on protostellar luminosity studies that have developed a relationship that correlates a YSO's mid-infrared luminosity to its bolometric luminosity \citep[see section 3.2 of][]{Kryukova12}. The mid-infrared flux is defined as a sum of 2MASS J,H, and K fluxes, IRAC 1,2,3 and 4 fluxes and the MIPS 24$\mu$m flux with coefficients determined by Equation (6) of \citet{Kryukova12}. We use our infrared source extractions to create mid-infrared fluxes for three of our protostellar sources and estimate their bolometric luminosities using Equation (5) from \citet{Dunham13}. We re-calculated our protostars' infrared spectral index, $\alpha_{\text{IR}}$, using the IRAC 3.6 $\mu$m band and the MIPS 24 $\mu$m band to be consistent with the derivation found in the former study.

\autoref{fig:radio_corr} shows three of our protostellar sources overlaid on the 3.6 cm luminosity and 6.0 cm luminosity vs. bolometric luminosity correlations provided by \citet{Shirley07}. Using this sample, \citet{Shirley07} report correlations of 
\begin{align*}
\log(L_{3.6}/L_{\text{1 mJy kpc$^{2}$}}) = &-(2.24 \pm 0.03)\\
&+ (0.71 \pm 0.01)\log(L_{\text{bol}}/L_{\odot}),\\[.1in]
\log(L_{6.0}/L_{\text{1 mJy kpc$^{2}$}}) = &-(2.51 \pm 0.03)\\
&+ (0.87 \pm 0.02)\log(L_{\text{bol}}/L_{\odot})
\end{align*}
for the 3.6 cm and 6.0 cm emission components respectively. Because the detection and characterization of low luminosity protostars is still an ongoing process, it is hard to tell for sure what the trend of this relationship is at the low-luminosity end where our protostars reside. However, deep radio studies by \citet{Scaife11b} suggest that this relationship may tentatively extend linearly from log bolometric luminosities of 1 down to -1. To compare this relationship to our sample of protostars, we used our derived radio spectral indices to extrapolate our 4.1 cm and 6.3 cm radio fluxes to 3.6 and 6.0 radio fluxes. The radio luminosities use an assumed distance of 429 pc to Serpens South. Errors on our derived bolometric luminosity account for errors in our infrared fluxes, infrared spectral indices, and errors in the scaling relationships provided by \citet{Kryukova12} and \citet{Dunham13}. Within our errors, we find that our data roughly agrees with the relationship as it currently stands. Although visually our data might seem to create some tension at the low-luminosity end, this is complicated by the fact that this $S_{\text{radio}}$ vs $L_{\text{bol}}$ relationship does not presently include all radio detections of protostars, has an intrinsic scatter, can be affected by factors such as emission variability and has yet to be fully probed at the low-luminosity end.

Our derived mid-infrared luminosities could be affected by the fact that we do not have 2MASS J, H, K band extractions for some of our protostars (\autoref{fig:infra_SEDs}). To test how sensitive our derived luminosities are to this lack of data, we artificially added JHK fluxes that roughly matched an extrapolation of the protostars' SED and found that the newly derived bolometric luminosities change by at most 0.03 in log space: much smaller than the errors already associated with the bolometric luminosity derivation. We therefore conclude that we are largely unaffected by the fact that we do not have J, H, K fluxes for determining our protostars' mid-infrared luminosity.

\section{Conclusions}
\label{sec:conclusions}
We have presented deep, interferometric radio observations of centimeter continuum emission from the core of Serpens South with the Karl G. Jansky Very Large Array. We detect radio emission from a number of YSO candidates identified in the infrared and millimeter and from one previously unidentified source. For Class 0 protostars identified in the millimeter but not detected in the infrared, we provide definitive evidence for a compact central source through our detection of thermal radio emission, indicative of outflows expected during the protostellar phase. Our 3 arcsecond resolution also allows us to separate and resolve individual YSO sources that were previously confused by lower spatial resolution studies. Out of the 40+ infrared YSO candidates in our field of view, we detect radio emission at or above a level of 50 $\mu$Jy/beam for only six. We also detect emission at a weaker level for another six that lie within 2 arcseconds of infrared YSO candidates with IR detections in some but not all of the IRAC 1, 2, 3, 4 and MIPS 24$\mu$m band, providing motivation for a higher sensitivity follow-up study to detect more embedded objects in Serpens South. Our detection of YSOs overlooked by infrared studies reinforce the notion that longer wavelength studies at millimeter and radio wavelengths are necessary to construct accurate demographics of prestellar populations. Finally, the fact that \citet{OrtizLeon15} did not make any firm detections of our YSO candidates suggests that either (1) the sources have significant time variability in their emission and/or (2) that the high-angular resolution of \citet{OrtizLeon15} ($\sim0.3$arcsec) over-resolved the emission structure from these YSOs. The latter, if true, would suggest that there is a significant source of thermal emission originating at or above size scales of hundreds of AU.

Although we cannot probe the inner regions of the collimated outflows that drive YSO thermal radio emission, our detections provide future high resolution studies a starting point for where thermal radio emission exists and its approximate strength. However, in fitting two-component Gaussians to our radio sources with CASA'a IMFIT task, we recover a deconvolved beam size for VLA 11 in our 7.25 GHz image that agrees with a previous work that found a jet emanating from VLA 11 in roughly the same direction \citep{Teixeira12}. Serpens South has also been found to drive large scale molecular outflows \citep{Nakamura11,Plunkett15}. Our strongest source of thermal radio emission is VLA 13, which lies almost directly in the center of both Serpens South's core and the peak of its extended sub-millimeter emission. We therefore speculate that VLA 13 could be a dominate driver of molecular outflow in the region.

We cross reference infrared catalogues and run source extractions over near and far infrared images from 2 $\mu$m to 70 $\mu$m to build infrared SEDs, which we use to both classify our protostellar sources and derive mid-infrared \& bolometric luminosities. We find relative agreement with the known radio vs. bolometric luminosity relationship of protostars for three of our sources given our errors. 

Although the core of Serpens South may seem to be relatively radio quiet judging from the high-resolution radio survey of \citet{OrtizLeon15}, we find a host of interesting radio sources that show clear signs of being protostellar in nature that  have for the first time shed light on the radio properties of YSOs in the core of the Serpens South infrared dark cloud. Future radio studies of Serpens South would benefit from multi-epoch, polarization-capable studies that have an angular resolution of $\sim$1 arcsecond and a point source sensitivity of less than 10 $\mu$Jy/beam in order to detect a larger number of young stellar objects.

\

The observations for this paper were conducted as part of the NRAO REU program funded by the National Science Foundation. N.S.K., J.A.K., and A.M. were summer students at the NRAO when the observations utilized by this study were taken. The National Radio Astronomy Observatory is a facility of the National Science Foundation operated under cooperative agreement by Associated Universities, Inc. J.J.T. acknowledges support provided by NASA through Hubble Fellowship grant \#HST-HF-51300.0 A awarded by the Space Telescope Science Institute, which is operated by the Association of Universities for Research in Astronomy, Inc., for NASA, under contract NAS 5-26555. This work is supported by grant 639.041.439 from the Netherlands Organisation for Scientific Research (NWO).

\bibliographystyle{apj}		
\bibliography{paper}

\begin{thebibliography}{}
\expandafter\ifx\csname natexlab\endcsname\relax\def\natexlab#1{#1}\fi

\bibitem[{{AMI Consortium} {et~al.}(2011{\natexlab{a}}){AMI Consortium},
  {Scaife}, {Hatchell}, {Davies}, {Franzen}, {Grainge}, {Hobson},
  {Hurley-Walker}, {Lasenby}, {Olamaie}, {Perrott}, {Pooley},
  {Rodr{\'{\i}}guez-Gonz{\'a}lvez}, {Saunders}, {Schammel}, {Scott},
  {Shimwell}, {Titterington}, \& {Waldram}}]{Scaife11a}
{AMI Consortium}, {Scaife}, A.~M.~M., {Hatchell}, J., {et~al.}
  2011{\natexlab{a}}, \mnras, 415, 893

\bibitem[{{AMI Consortium} {et~al.}(2011{\natexlab{b}}){AMI Consortium},
  {Scaife}, {Curtis}, {Davies}, {Franzen}, {Grainge}, {Hobson},
  {Hurley-Walker}, {Lasenby}, {Olamaie}, {Pooley},
  {Rodr{\'{\i}}guez-Gonz{\'a}lvez}, {Saunders}, {Schammel}, {Scott},
  {Shimwell}, {Titterington}, {Waldram}, \& {Zwart}}]{Scaife11b}
{AMI Consortium}, {Scaife}, A.~M.~M., {Curtis}, E.~I., {et~al.}
  2011{\natexlab{b}}, \mnras, 410, 2662

\bibitem[{{Andre} {et~al.}(2000){Andre}, {Ward-Thompson}, \&
  {Barsony}}]{Andre00}
{Andre}, P., {Ward-Thompson}, D., \& {Barsony}, M. 2000, Protostars and Planets
  IV, 59

\bibitem[{{Anglada}(1995)}]{Anglada95}
{Anglada}, G. 1995, in Revista Mexicana de Astronomia y Astrofisica, vol. 27,
  Vol.~1, Revista Mexicana de Astronomia y Astrofisica Conference Series, ed.
  S.~{Lizano} \& J.~M. {Torrelles}, 67

\bibitem[{{Anglada} {et~al.}(1998){Anglada}, {Villuendas}, {Estalella},
  {Beltr{\'a}n}, {Rodr{\'{\i}}guez}, {Torrelles}, \& {Curiel}}]{Anglada98}
{Anglada}, G., {Villuendas}, E., {Estalella}, R., {et~al.} 1998, \aj, 116, 2953

\bibitem[{{Beltr{\'a}n} {et~al.}(2001){Beltr{\'a}n}, {Estalella}, {Anglada},
  {Rodr{\'{\i}}guez}, \& {Torrelles}}]{Beltran01}
{Beltr{\'a}n}, M.~T., {Estalella}, R., {Anglada}, G., {Rodr{\'{\i}}guez},
  L.~F., \& {Torrelles}, J.~M. 2001, \aj, 121, 1556

\bibitem[{{Bontemps} {et~al.}(2010){Bontemps}, {Andr{\'e}}, {K{\"o}nyves},
  {Men'shchikov}, {Schneider}, {Maury}, {Peretto}, {Arzoumanian}, {Attard},
  {Motte}, {Minier}, {Didelon}, {Saraceno}, {Abergel}, {Baluteau}, {Bernard},
  {Cambr{\'e}sy}, {Cox}, {di Francesco}, {di Giorgo}, {Griffin}, {Hargrave},
  {Huang}, {Kirk}, {Li}, {Martin}, {Mer{\'{\i}}n}, {Molinari}, {Olofsson},
  {Pezzuto}, {Prusti}, {Roussel}, {Russeil}, {Sauvage}, {Sibthorpe},
  {Spinoglio}, {Testi}, {Vavrek}, {Ward-Thompson}, {White}, {Wilson},
  {Woodcraft}, \& {Zavagno}}]{Bontemps10}
{Bontemps}, S., {Andr{\'e}}, P., {K{\"o}nyves}, V., {et~al.} 2010, \aap, 518,
  L85

\bibitem[{{Briggs}(1995)}]{Briggs95}
{Briggs}, D.~S. 1995, dissertation, New Mexico Inst. of Mining and Technology

\bibitem[{{Clark}(1980)}]{Clark80}
{Clark}, B.~G. 1980, \aap, 89, 377

\bibitem[{{Crapsi} {et~al.}(2008){Crapsi}, {van Dishoeck}, {Hogerheijde},
  {Pontoppidan}, \& {Dullemond}}]{Crapsi08}
{Crapsi}, A., {van Dishoeck}, E.~F., {Hogerheijde}, M.~R., {Pontoppidan},
  K.~M., \& {Dullemond}, C.~P. 2008, \aap, 486, 245

\bibitem[{{Curiel} {et~al.}(1989){Curiel}, {Rodriguez}, {Bohigas}, {Roth},
  {Canto}, \& {Torrelles}}]{Curiel89}
{Curiel}, S., {Rodriguez}, L.~F., {Bohigas}, J., {et~al.} 1989, Astrophysical
  Letters and Communications, 27, 299

\bibitem[{{Dunham} {et~al.}(2013){Dunham}, {Arce}, {Allen}, {Evans},
  {Broekhoven-Fiene}, {Chapman}, {Cieza}, {Gutermuth}, {Harvey}, {Hatchell},
  {Huard}, {Kirk}, {Matthews}, {Mer{\'{\i}}n}, {Miller}, {Peterson}, \&
  {Spezzi}}]{Dunham13}
{Dunham}, M.~M., {Arce}, H.~G., {Allen}, L.~E., {et~al.} 2013, \aj, 145, 94

\bibitem[{{Dunham} {et~al.}(2015){Dunham}, {Allen}, {Evans},
  {Broekhoven-Fiene}, {Cieza}, {Di Francesco}, {Gutermuth}, {Harvey},
  {Hatchell}, {Heiderman}, {Huard}, {Johnstone}, {Kirk}, {Matthews}, {Miller},
  {Peterson}, \& {Young}}]{Dunham15}
{Dunham}, M.~M., {Allen}, L.~E., {Evans}, II, N.~J., {et~al.} 2015, ArXiv
  e-prints, arXiv:1508.03199

\bibitem[{{Dzib} {et~al.}(2010){Dzib}, {Loinard}, {Mioduszewski}, {Boden},
  {Rodr{\'{\i}}guez}, \& {Torres}}]{Dzib10}
{Dzib}, S., {Loinard}, L., {Mioduszewski}, A.~J., {et~al.} 2010, \apj, 718, 610

\bibitem[{{Dzib} {et~al.}(2011){Dzib}, {Loinard}, {Mioduszewski}, {Boden},
  {Rodr{\'{\i}}guez}, \& {Torres}}]{Dzib11}
{Dzib}, S., {Loinard}, L., {Mioduszewski}, A.~J., {et~al.} 2011, in Revista
  Mexicana de Astronomia y Astrofisica Conference Series, Vol.~40, Revista
  Mexicana de Astronomia y Astrofisica Conference Series, 231--232

\bibitem[{{Dzib} {et~al.}(2014){Dzib}, {Loinard}, {Rodr{\'{\i}}guez}, \&
  {Galli}}]{Dzib14a}
{Dzib}, S.~A., {Loinard}, L., {Rodr{\'{\i}}guez}, L.~F., \& {Galli}, P. 2014,
  \apj, 788, 162

\bibitem[{{Dzib} {et~al.}(2013){Dzib}, {Loinard}, {Mioduszewski},
  {Rodr{\'{\i}}guez}, {Ortiz-Le{\'o}n}, {Pech}, {Rivera}, {Torres}, {Boden},
  {Hartmann}, {Evans}, {Brice{\~n}o}, \& {Tobin}}]{Dzib13}
{Dzib}, S.~A., {Loinard}, L., {Mioduszewski}, A.~J., {et~al.} 2013, \apj, 775,
  63

\bibitem[{{Evans} {et~al.}(2009){Evans}, {Dunham}, {J{\o}rgensen}, {Enoch},
  {Mer{\'{\i}}n}, {van Dishoeck}, {Alcal{\'a}}, {Myers}, {Stapelfeldt},
  {Huard}, {Allen}, {Harvey}, {van Kempen}, {Blake}, {Koerner}, {Mundy},
  {Padgett}, \& {Sargent}}]{Evans09}
{Evans}, II, N.~J., {Dunham}, M.~M., {J{\o}rgensen}, J.~K., {et~al.} 2009,
  \apjs, 181, 321

\bibitem[{{Feigelson} \& {Montmerle}(1999)}]{Feigelson99}
{Feigelson}, E.~D., \& {Montmerle}, T. 1999, \araa, 37, 363

\bibitem[{{Fern{\'a}ndez-L{\'o}pez} {et~al.}(2014){Fern{\'a}ndez-L{\'o}pez},
  {Arce}, {Looney}, {Mundy}, {Storm}, {Teuben}, {Lee}, {Segura-Cox}, {Isella},
  {Tobin}, {Rosolowsky}, {Plunkett}, {Kwon}, {Kauffmann}, {Ostriker}, {Tassis},
  {Shirley}, \& {Pound}}]{Fernandez-Lopez14}
{Fern{\'a}ndez-L{\'o}pez}, M., {Arce}, H.~G., {Looney}, L., {et~al.} 2014,
  \apjl, 790, L19

\bibitem[{{Fomalont} {et~al.}(1991){Fomalont}, {Windhorst}, {Kristian}, \&
  {Kellerman}}]{Fomalont91}
{Fomalont}, E.~B., {Windhorst}, R.~A., {Kristian}, J.~A., \& {Kellerman}, K.~I.
  1991, \aj, 102, 1258

\bibitem[{{Friesen} {et~al.}(2013){Friesen}, {Medeiros}, {Schnee}, {Bourke},
  {Francesco}, {Gutermuth}, \& {Myers}}]{Friesen13}
{Friesen}, R.~K., {Medeiros}, L., {Schnee}, S., {et~al.} 2013, \mnras, 436,
  1513

\bibitem[{{Ghavamian} \& {Hartigan}(1998)}]{Ghavamian98}
{Ghavamian}, P., \& {Hartigan}, P. 1998, \apj, 501, 687

\bibitem[{{Greene} {et~al.}(1994){Greene}, {Wilking}, {Andre}, {Young}, \&
  {Lada}}]{Greene94}
{Greene}, T.~P., {Wilking}, B.~A., {Andre}, P., {Young}, E.~T., \& {Lada},
  C.~J. 1994, \apj, 434, 614

\bibitem[{{Gutermuth} {et~al.}(2009){Gutermuth}, {Megeath}, {Myers}, {Allen},
  {Pipher}, \& {Fazio}}]{Gutermuth09}
{Gutermuth}, R.~A., {Megeath}, S.~T., {Myers}, P.~C., {et~al.} 2009, \apjs,
  184, 18

\bibitem[{{Gutermuth} {et~al.}(2008){Gutermuth}, {Bourke}, {Allen}, {Myers},
  {Megeath}, {Matthews}, {J{\o}rgensen}, {Di Francesco}, {Ward-Thompson},
  {Huard}, {Brooke}, {Dunham}, {Cieza}, {Harvey}, \& {Chapman}}]{Gutermuth08}
{Gutermuth}, R.~A., {Bourke}, T.~L., {Allen}, L.~E., {et~al.} 2008, \apjl, 673,
  L151

\bibitem[{{H{\"o}gbom}(1974)}]{Hogbom74}
{H{\"o}gbom}, J.~A. 1974, \aaps, 15, 417

\bibitem[{{Kirk} {et~al.}(2013){Kirk}, {Myers}, {Bourke}, {Gutermuth},
  {Hedden}, \& {Wilson}}]{Kirk13}
{Kirk}, H., {Myers}, P.~C., {Bourke}, T.~L., {et~al.} 2013, \apj, 766, 115

\bibitem[{{Konyves} {et~al.}(2015){Konyves}, {Andre}, {Men'shchikov},
  {Palmeirim}, {Arzoumanian}, {Schneider}, {Roy}, {Didelon}, {Maury},
  {Shimajiri}, {Di Francesco}, {Bontemps}, {Peretto}, {Benedettini}, {Bernard},
  {Elia}, {Griffin}, {Hill}, {Kirk}, {Ladjelate}, {Marsh}, {Martin}, {Motte},
  {Nguyen Luong}, {Pezzuto}, {Roussel}, {Rygl}, {Sadavoy}, {Schisano},
  {Spinoglio}, {Ward-Thompson}, \& {White}}]{Konyves15}
{Konyves}, V., {Andre}, P., {Men'shchikov}, A., {et~al.} 2015, ArXiv e-prints,
  arXiv:1507.05926

\bibitem[{{Kryukova} {et~al.}(2012){Kryukova}, {Megeath}, {Gutermuth},
  {Pipher}, {Allen}, {Allen}, {Myers}, \& {Muzerolle}}]{Kryukova12}
{Kryukova}, E., {Megeath}, S.~T., {Gutermuth}, R.~A., {et~al.} 2012, \aj, 144,
  31

\bibitem[{{Kuhn} {et~al.}(2010){Kuhn}, {Getman}, {Feigelson}, {Reipurth},
  {Rodney}, \& {Garmire}}]{Kuhn10}
{Kuhn}, M.~A., {Getman}, K.~V., {Feigelson}, E.~D., {et~al.} 2010, \apj, 725,
  2485

\bibitem[{{Maury} {et~al.}(2011){Maury}, {Andr{\'e}}, {Men'shchikov},
  {K{\"o}nyves}, \& {Bontemps}}]{Maury11}
{Maury}, A.~J., {Andr{\'e}}, P., {Men'shchikov}, A., {K{\"o}nyves}, V., \&
  {Bontemps}, S. 2011, \aap, 535, A77

\bibitem[{{Nakamura} \& {Li}(2014)}]{Nakamura14a}
{Nakamura}, F., \& {Li}, Z.-Y. 2014, \apj, 783, 115

\bibitem[{{Nakamura} {et~al.}(2011){Nakamura}, {Sugitani}, {Shimajiri},
  {Tsukagoshi}, {Higuchi}, {Nishiyama}, {Kawabe}, {Takami}, {Karr},
  {Gutermuth}, \& {Wilson}}]{Nakamura11}
{Nakamura}, F., {Sugitani}, K., {Shimajiri}, Y., {et~al.} 2011, \apj, 737, 56

\bibitem[{{Ortiz-Le{\'o}n} {et~al.}(2015){Ortiz-Le{\'o}n}, {Loinard},
  {Mioduszewski}, {Dzib}, {Rodr{\'{\i}}guez}, {Pech}, {Rivera}, {Torres},
  {Boden}, {Hartmann}, {Evans}, {Brice{\~n}o}, {Tobin}, {Kounkel}, \&
  {Gonz{\'a}lez-L{\'o}pezlira}}]{OrtizLeon15}
{Ortiz-Le{\'o}n}, G.~N., {Loinard}, L., {Mioduszewski}, A.~J., {et~al.} 2015,
  \apj, 805, 9

\bibitem[{{Plunkett} {et~al.}(2015{\natexlab{a}}){Plunkett}, {Arce}, {Corder},
  {Dunham}, {Garay}, \& {Mardones}}]{Plunkett15}
{Plunkett}, A.~L., {Arce}, H.~G., {Corder}, S.~A., {et~al.} 2015{\natexlab{a}},
  \apj, 803, 22

\bibitem[{{Plunkett} {et~al.}(2015{\natexlab{b}}){Plunkett}, {Arce},
  {Mardones}, {van Dokkum}, {Dunham}, {Fern{\'a}ndez-L{\'o}pez}, {Gallardo}, \&
  {Corder}}]{Plunkett15b}
{Plunkett}, A.~L., {Arce}, H.~G., {Mardones}, D., {et~al.} 2015{\natexlab{b}},
  \nat, 527, 70

\bibitem[{{Rodriguez} {et~al.}(1989){Rodriguez}, {Myers}, {Cruz-Gonzalez}, \&
  {Terebey}}]{Rodriguez89}
{Rodriguez}, L.~F., {Myers}, P.~C., {Cruz-Gonzalez}, I., \& {Terebey}, S. 1989,
  \apj, 347, 461

\bibitem[{{Rodr{\'{\i}}guez} {et~al.}(2003){Rodr{\'{\i}}guez}, {Porras},
  {Claussen}, {Curiel}, {Wilner}, \& {Ho}}]{Rodriguez03}
{Rodr{\'{\i}}guez}, L.~F., {Porras}, A., {Claussen}, M.~J., {et~al.} 2003,
  \apjl, 586, L137

\bibitem[{{Rodr{\'{\i}}guez} {et~al.}(2010){Rodr{\'{\i}}guez}, {Rodney}, \&
  {Reipurth}}]{Rodriguez10}
{Rodr{\'{\i}}guez}, L.~F., {Rodney}, S.~A., \& {Reipurth}, B. 2010, \aj, 140,
  968

\bibitem[{{Schnee} {et~al.}(2012){Schnee}, {Di Francesco}, {Enoch}, {Friesen},
  {Johnstone}, \& {Sadavoy}}]{Schnee12}
{Schnee}, S., {Di Francesco}, J., {Enoch}, M., {et~al.} 2012, \apj, 745, 18

\bibitem[{{Schwab}(1984)}]{Schwab84}
{Schwab}, F.~R. 1984, \aj, 89, 1076

\bibitem[{{Shirley} {et~al.}(2007){Shirley}, {Claussen}, {Bourke}, {Young}, \&
  {Blake}}]{Shirley07}
{Shirley}, Y.~L., {Claussen}, M.~J., {Bourke}, T.~L., {Young}, C.~H., \&
  {Blake}, G.~A. 2007, \apj, 667, 329

\bibitem[{{Strai{\v z}ys} {et~al.}(2003){Strai{\v z}ys}, {{\v C}ernis}, \&
  {Barta{\v s}i{\= u}t{\.e}}}]{Straizys03}
{Strai{\v z}ys}, V., {{\v C}ernis}, K., \& {Barta{\v s}i{\= u}t{\.e}}, S. 2003,
  \aap, 405, 585

\bibitem[{{Sugitani} {et~al.}(2011){Sugitani}, {Nakamura}, {Watanabe},
  {Tamura}, {Nishiyama}, {Nagayama}, {Kandori}, {Nagata}, {Sato}, {Gutermuth},
  {Wilson}, \& {Kawabe}}]{Sugitani11}
{Sugitani}, K., {Nakamura}, F., {Watanabe}, M., {et~al.} 2011, \apj, 734, 63

\bibitem[{{Tanaka} {et~al.}(2013){Tanaka}, {Nakamura}, {Awazu}, {Shimajiri},
  {Sugitani}, {Onishi}, {Kawabe}, {Yoshida}, \& {Higuchi}}]{Tanaka13}
{Tanaka}, T., {Nakamura}, F., {Awazu}, Y., {et~al.} 2013, \apj, 778, 34

\bibitem[{{Teixeira} {et~al.}(2012){Teixeira}, {Kumar}, {Bachiller}, \&
  {Grave}}]{Teixeira12}
{Teixeira}, G.~D.~C., {Kumar}, M.~S.~N., {Bachiller}, R., \& {Grave}, J.~M.~C.
  2012, \aap, 543, A51

\bibitem[{{Wu} {et~al.}(2004){Wu}, {Wei}, {Zhao}, {Shi}, {Yu}, {Qin}, \&
  {Huang}}]{Wu04}
{Wu}, Y., {Wei}, Y., {Zhao}, M., {et~al.} 2004, \aap, 426, 503

\end{thebibliography}

\end{document}